\documentclass[prb,pre,aps,twocolumn,amsmath,amssymb,floatfix,
superscriptaddress]{revtex4-1}
 \DeclareUnicodeCharacter{0301}{\hspace{-1ex}\'{ }}
\usepackage[dvips]{graphics}
\usepackage{graphicx}
\usepackage{color}
\usepackage{blindtext}
\definecolor{dred}{rgb}{0.75,0,0}
\usepackage{graphicx}
\usepackage{dcolumn}
\usepackage{bm}
\usepackage{xcolor}
\usepackage{braket}
\usepackage{physics}
\usepackage{appendix}
\usepackage{hyperref}
\hypersetup{
 colorlinks=true,
citecolor=magenta,filecolor=blue,
 linkcolor=magenta,
 }

\usepackage{graphicx}
\usepackage{dcolumn}
\usepackage{bm}
\usepackage{xcolor}

\definecolor{codegreen}{rgb}{0,0.6,0}
\definecolor{codegray}{rgb}{0.5,0.5,0.5}
\definecolor{codepurple}{rgb}{0.58,0,0.82}
\definecolor{backcolour}{rgb}{0.95,0.95,0.92}

\begin{document}

\preprint{APS/123-QED}

\title{\textcolor{purple}{Designer quantum states on a fractal substrate: compact localization, flat bands and the edge modes}} 

\author{Sougata Biswas}
\affiliation{Department of Physics, Presidency University, 86/1 College Street, Kolkata, West Bengal - 700 073, India}
\email{biswassougata3@gmail.com}
\author{Arunava Chakrabarti}
\affiliation{Department of Physics, Presidency University, 86/1 College Street, Kolkata, West Bengal - 700 073, India}
\email{arunava.physics@presiuniv.ac.in}
\date{\today}

\begin{abstract}
Compact localized single particle eigenstates on a deterministic fractal substrate, modelled by a triangular Sierpinski gasket of arbitrarily large size, are unravelled and examined analytically. We prescribe an exact real space renormalization group (RSRG) decimation scheme within a tight binding formalism to discern these states, and argue that the number of such states can be infinite if the fractal substrate is enlarged to its thermodynamic limit. Interestingly, these localized states turn out to populate the non-dispersive, flat bands in a periodic array of Sierpinski gasket motifs, however large they may be. Our results match and corroborate the recently observed compact localized, flat band states engineered on two dimensional photonic waveguide networks with a fractal geometry, and provide a whole subset of them, which, in principle, should be observable in fractal photonic lattice experiments. 
\end{abstract}

\maketitle

\section{Introduction}
\label{intro}
Lattice models of quantum networks, artificially tailored, and exhibiting translational invariance, have drawn immense attention of the condensed matter physics community in recent times. One main reason for this is the occurrence of non-dispersive, flat Bloch bands that result out of a destructive quantum interference, triggered by the local environment around a site or a cluster of sites in such lattices~\cite{leykam,derzhko1,bergholtz}. The group velocity of an excitation occupying a flat band (FB) becomes zero, resulting in an infinitely large efective mass of the excitation. Consequently, the wave transmission is suppressed. It's a localization of waves that does not need a disorder in the system.

In a system exhibiting flat bands (FB's) typically the amplitude of the wavefunction is non-zero over a cluster of sites, an `island' if we may call it so, that is separated from other such islands by sites where the amplitude of the wave function is zero. Such states are macroscopically degenerate, and are termed {\it compact localized states} (CLS). The CLS have recently been receiving serious attention as precise engineering of the wave function of an excitation is now a days achievable in networks on an atomic scale. Creation of quantum nanostructures with well defined geometries, using an atom-by-atom manipulation and the unprecedented  tunability acheived on an atomic scale are opening up huge opportunities to explore, not only the fundamental physics of the quantum world, but the potential of such designer states in transmitting quantum information~\cite{immanuel,blatt,rodrigo1,rojas,xia} as well.

One of the first theoretical works, predicting the existence of FB's or the CLS dates back to the late 1980's~\cite{sutherland}, though the fabrication technologies needed to explore and expose the FB's have only very recently been achieved to realize artificial flat band lattices. For example, experiments involving optical systems and ultracold atoms, have been instrumental~\cite{gersen,seba,rodrigo2,zong,travkin} in `seeing' the flat bands in some designer lattices. The studies have subsequently been substantiated by fabrications of artificial materials through `atom manipulation' using the scanning tunneling microscope~\cite{drost,slot,gardenier,alexander}.

Certain network models with inherent translational invariance, described within a tight binding formalism, provide a useful platform to learn and analyze the existence of FB's and the CLS~\cite{miyahara,mati,morales}. The importance of the  local symmetries of a discrete Hamiltonian in generating the CLS has been appreciated, and explored~\cite{rontgen}. 
\begin{figure*}[ht]
\centering
(a)\includegraphics[width=.85\columnwidth]{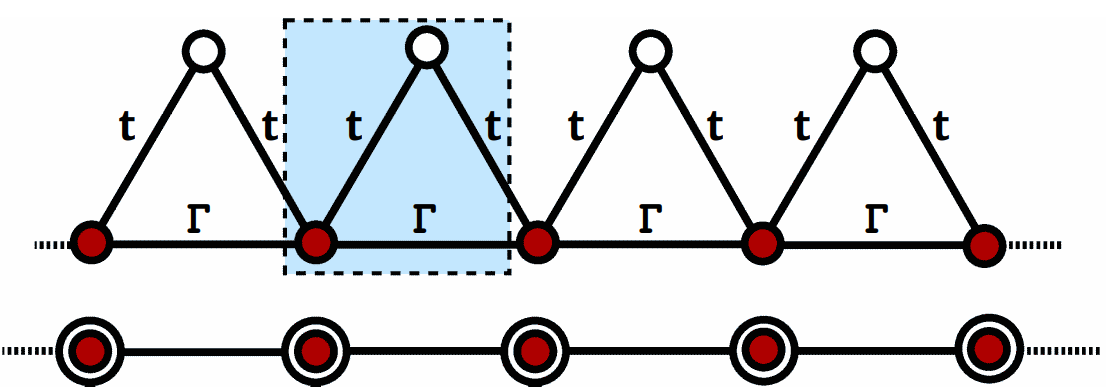}
(b)\includegraphics[width=.65\columnwidth]{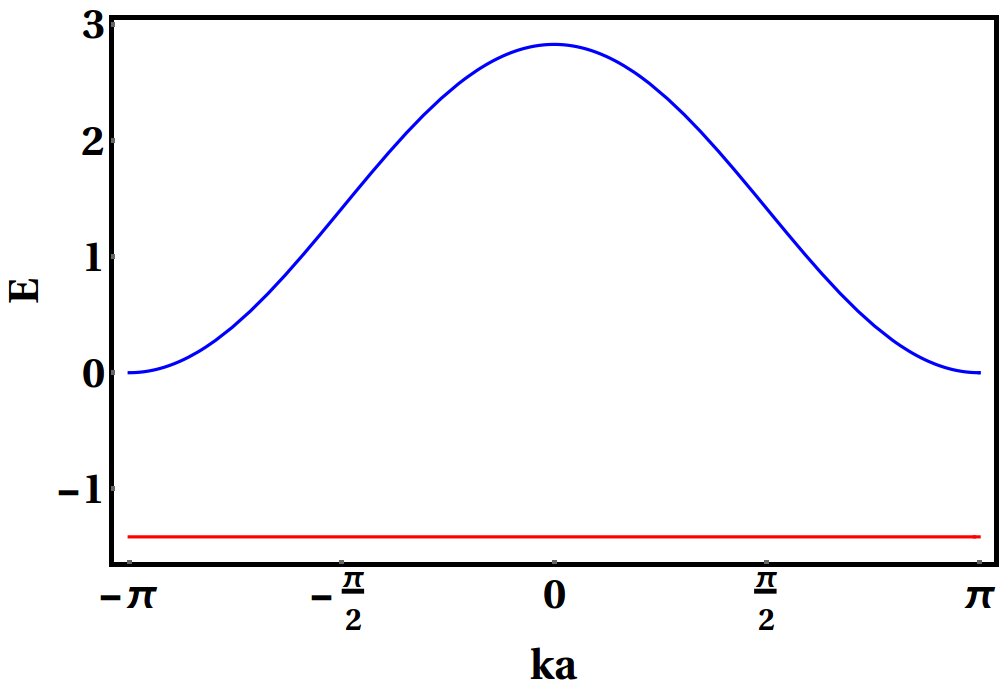}
(c)\includegraphics[width=.85\columnwidth]{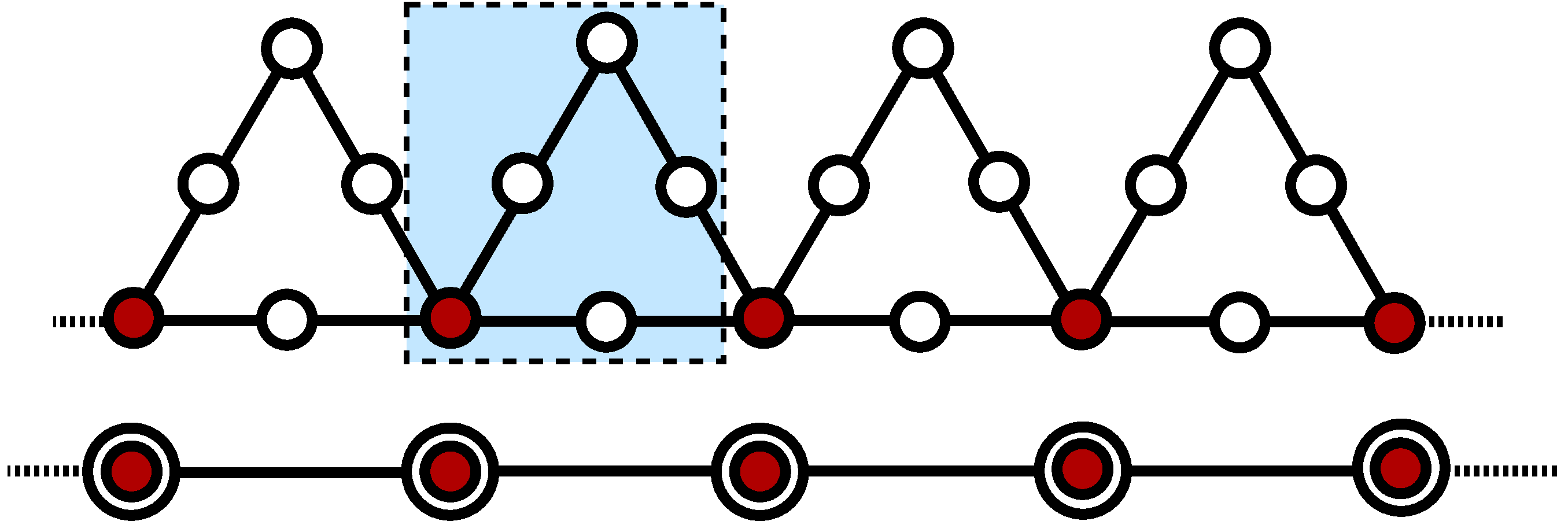}
(d)\includegraphics[width=.65\columnwidth]{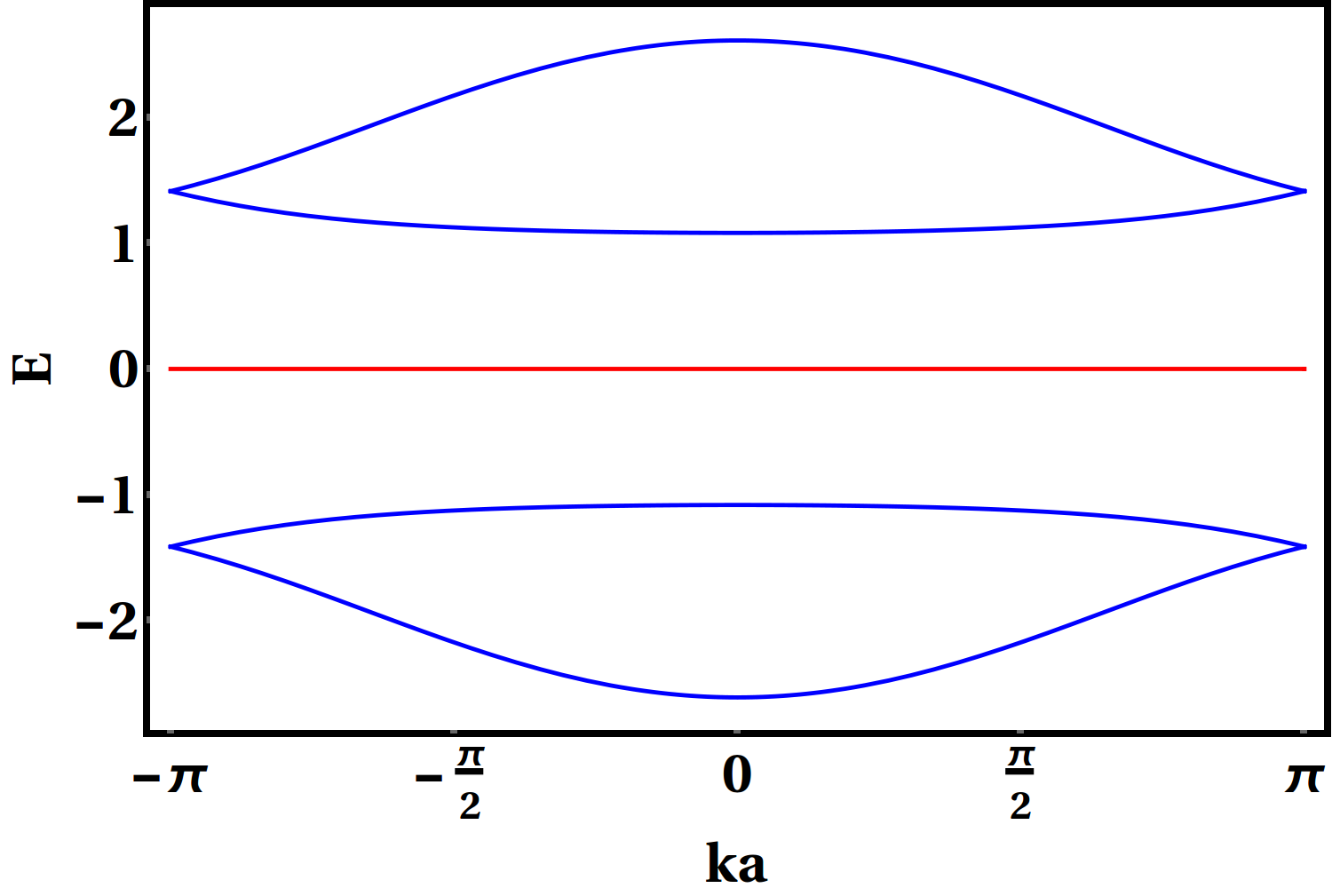}
(e)\includegraphics[width=.85\columnwidth]{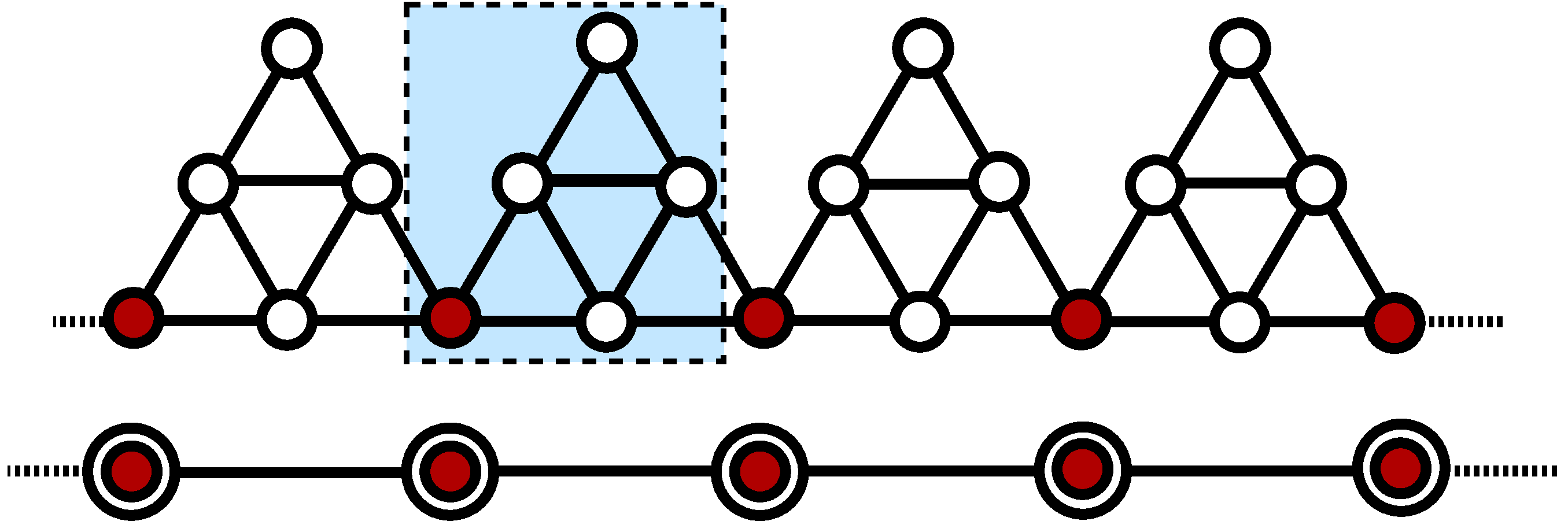}
(f)\includegraphics[width=.65\columnwidth]{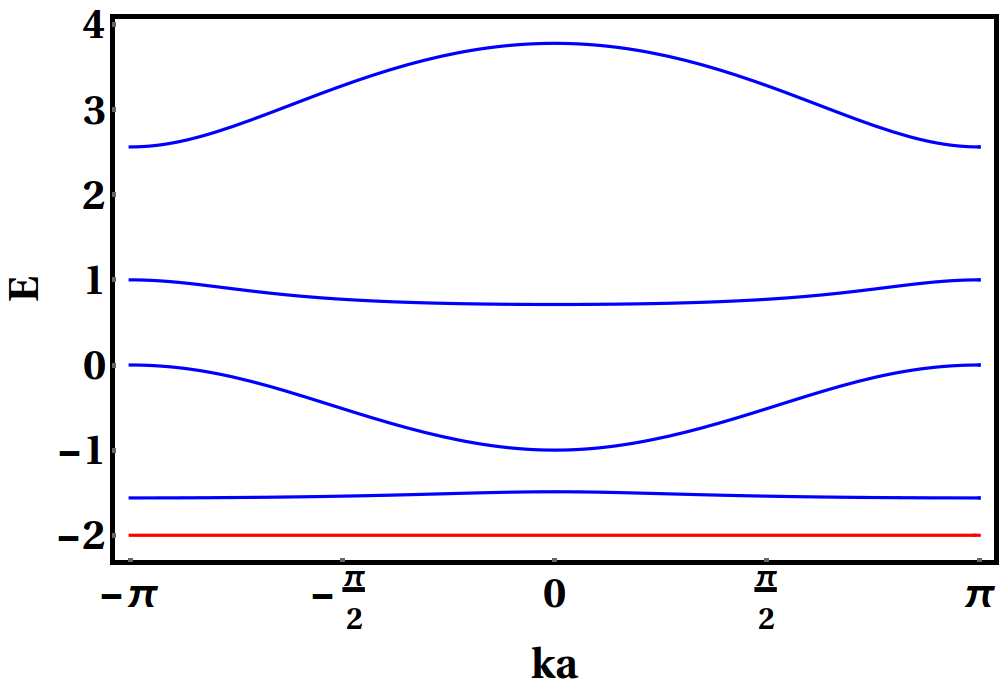}

\caption{(Color online) (a) A simple periodic sawtooth array with two different types of hopping integrals $\Gamma$ and $t$, (c) a periodic triangular array with one extra atomic site in each arm, and with a uniform hopping $t$, and (e) a periodic array of the $G_1$ Sierpinski gasket network with uniform hopping $t$ along each arm. Corresponding  Energy-wave vector dispersion relations are shown with $\epsilon = 0 $, $ t = 1 $, $\Gamma = 1/\sqrt{2}$ in (b), and with $\epsilon = 0 $ and  $t = 1$ in (d) and (f) respectively. Flat bands are drawn in red. The flat bands energies are, (b) $ E = -\sqrt{2} $ (non degenerate), (d) $E = 0$ (non degenerate), (f)$E = -2$ (non degenerate). }  
\label{fig-ek}
\end{figure*}

Unravelling the FB's and the CLS is generally (and conveniently) done by diagonalizing the Hamiltonian of the `unit cell' of the network in $k$-space. This works easily when the unit cell has simple geometry, but turns out to be a tedious job in cases where the unit cell has an intricate internal structure. This issue has become pertinent and needs to be addressed, in particular, after a series of fascinating experiments reported laboratory realizations of the classic Sierpinski gasket fractal (SGF)~\cite{shang,tait,sun} and observation of CLS~\cite{xie,hanafi,liu,kempkes,xu} and photonic Floquet topological phase~\cite{yang} on engineered SGF substrates. The latter experiments were done on SGF structures grafted either mainly by the femtosecond laser writing techniques, or designed with evanescently coupled helical waveguides, for example.  Parallel theoretical proposals and analyses of topological phase transitions and edge states in fractal space~\cite{shriya, fischer,fremling, manna, brezinska} came up, exploring the unique geometry of an SGF. This domain of research has opened up the possibility of looking back at the well understood physics and geometry of the SGF and its variants~\cite{domany,banavar, wang,arunava} from a new perspective. Apart from opening up possible new avenues for functional devices, these designer networks throw new light on the physics of topological matter and the theory of Anderson localization~\cite{anderson}.

In this article we revisit the SGF in the light of the designer quantum states, which will form the CLS on such a fractal substrate. The recursive growth of an SGF pattern with an inherent self-similarity naturally invites an implementation of a real space renormalization group (RSRG) decimation  method~\cite{domany,wang,arunava}, that we will eventually explore. We want to emphasize that, the use of the RSRG method, especially for structural units that grow deterministically following a well defined recursive algorithm, gives us the opportunity to get the CLS patterns and the eigenvalues with more ease and elegance over the hard task of diagonalizing the unit cell matrices that becomes more and more difficult to frame as the unit cell grows towards its thermodynamic limit. We lay down our scheme in the present communication below. 

First of all we elucidate the strength of the RSRG decimation scheme to obtain the dispersion relations for three variants of a sawtooth chain, and obtain the $E-k$ diagrams.  In the first example, that of a simple array of triangles, our results exactly match the existing results obtained through a direct diagonalization of the Hamiltonian in $k$-space~\cite{mati}.  The next example is an extended version of basic sawtooh lattice, unaddressed so far, to the best of our knowledge, and the third example discusses a periodic array of the first generation SGF palquette, acting as the 'unit cell'.  In each case a flat, non-dispersive energy band is seen to exist which corresponds to a CLS in the basic isolated cluster. The design of such CLS will be discussed in details, in respect of an SGF. Subsequently, the RSRG method will be used efficiently to construct the CLS on an SGF of any arbitrarily large generation, and to exactly evaluate the corresponding energy eigenvalues.  

Secondly, we construct periodic chains in one dimension by repeating unit cells that are SGF's with sequentially increasing generations. We use a decimation scheme to map the structure onto a purely one dimensional geometry with energy dependent on-site potentials and the hopping integrals, and find that the $E-k$ spectrum displays both the dispersive and non-dispervive energy bands that get more and more densely packed as the structural motif contains higher and higher generation SGF units. The flat, non-dispersive bands occur exactly at the same energy eigenvalues for which the CLS are formed on the corresponding finite generation SGF substrate. A plethora of such CLS evolves as the SGF motif grows in size.

Before ending the introduction it is worth mentioning that, 
the periodic array of the SGF units essentially provides very interesting variants of networks in the `sawtooth' class. Sawtooth networks have drawn considerable attention in recent times as examples of geometrically frustrated lattices and have been the testing ground for studies of correlated systems~\cite{derzhko2}, adiabatic pumping~\cite{schulze}, quantum dynamics and Bloch-Zener oscillations~\cite{cai}, non-linear localized modes~\cite{magnus}, tunability of the bands using ultracold atomic platform~\cite{boong}, or, for discerning the CLS and flat bands with corresponding transport~\cite{maimaiti} or even the Josephson effect~\cite{peotta}. The SGF arrays thus open up an avenue to address these issues afresh, from the perspective of a designer quantum network that has microscopic details in its growth pattern  in a scale invariant unit cell.

In section II we explain the RSRG decimation methodology to obtain the dispersion relation for three quasi-one dimensional designer lattices. Section III deals with finite SGF's and how to recursively construct the CLS. The edge states are drawn on finite versions of the SGF and the band diagrams are shown explicitly, where the dispersive and the non-dispersive bands get closer and closer as the generation is increased. In section IV we draw conclusion.

\vspace{0.2cm}
\section{The $E-k$ dispersion relations for the sawtooth variants: A decimation scheme}

We refer to Fig.~\ref{fig-ek}, where a basic sawtooth array (BSA)(in (a)), a sawtooth superlattice (SSL) with just an extra atomic site on each side of the triangle (in (c)) and an SGF-sawtooth (SGFS) array with a first generation SGF~\cite{domany} embedded inside the triangle (in (e)) are shown. The `unit cell' for each lattice is shown in blue shade. 

We consider spinless, non-interacting Fermions on all these (and subsequent ones in other sections) systems and use the tight binding Hamiltonian, 
\begin{equation}
H = \epsilon \ket{i}\bra{i} + \sum_{<ij>} \tau_{ij} \ket{i}\bra{j} + h.c.
\label{ham}
\end{equation}
$\epsilon$ is the on-site potential (assumed constant), and $\tau_{ij}$ symbolizes the nearest neighbor hopping integral. For Fig.~\ref{fig-ek}(a) $\tau_{ij}$ is chosen to have two different values, viz, $t$ and $\Gamma$ corresponding to the particle's `hop' along the slanting arms and the horizonal arm of the triangle respectively, in order to match our results with the existing literature~\cite{mati}. In the rest of the figures and throughout this communication $\tau_{ij} = t$, a constant along every side of the embedded geometry.

Using the set of difference equations,
\begin{equation}
(E - \epsilon) \psi_i = \sum_{j} \tau_{ij} \psi_j
\label{diffeq}
\end{equation}
with $j$ representing the nearest neighbor sites of $i$, we decimate every `top' vertex (unfilled circle) in the BSA, shown in Fig.~\ref{fig-ek}(a), and retain the `base' atoms (red colored), which are now `renormalized' and lie on an effective one dimensional chain. The renormalized chain is shown just below the sawtooth array with encircled red sites, having an effective, energy dependent on-site potential, and an energy dependent nearest neighbor hopping integral, given by, 
\begin{eqnarray}
\tilde{\epsilon} & = & \epsilon + \frac{2t^2}{E-\epsilon} \nonumber \\
\tilde{t} & = & \Gamma + \frac{t^2}{E-\epsilon}
\label{rg1a}
\end{eqnarray}
The dispersion relation for the effective $1$-d chain can then easily be written down as, 
\begin{equation}
E = \tilde{\epsilon} + 2 \tilde{t} \cos~ka
\label{disp}
\end{equation}
that expands, with a choice of $\Gamma=t/\sqrt{2}$,  into~\cite{mati}, 
\begin{equation}
\left [E - (\epsilon-\sqrt{2}t)\right ] \left [ E - \epsilon-2 \sqrt{2}t \cos^2{(ka/2)} \right ] = 0
\label{saw-1}
\end{equation}
Here, $a$ is the effective lattice constant on the 1-d chain. Eq.~\eqref{saw-1} clearly shows a FB at $E = \epsilon-\sqrt{2}t$ that matches exactly with $E = \mp \sqrt{2}$ for $\epsilon=0$ and $t=\pm 1$, as chosen before~\cite{mati}. The bands are shown in Fig.~\ref{fig-ek}(b). Needless to say, the dispersive band also shows an exact match with the literature.

Fig.~\ref{fig-ek}(c) shows an extension of the sawtooth geometry, an SSL. The presence of an additional atom in each arm now results in a richer band spectrum (Fig.~\ref{fig-ek}(d)) showing a flat band at the centre of the spectrum $E=\epsilon=0$. The empty circles are decimated out, and the resulting 1-d chain (with encircled red colored sites) has a dispersion relation, 
\begin{equation}
(E-\epsilon) \left [1 - \left ( \frac{4 t^2 \alpha + 2 t^4 + 2 (\alpha t^2 + t^4) \cos~ ka}{\delta} \right ) \right ] = 0
\label{saw-2}
\end{equation}
Here, $\alpha=(E-\epsilon)^2 - 2 t^2$, and $\delta= \alpha (E-\epsilon)^2$.
The FB at $E=\epsilon$ is obvious.

In the SGFS, shown in Fig.~\ref{fig-ek}(e), we depict a sawtooth array where a first generation SGF replaces every basic triangle in Fig.~\ref{fig-ek}(a). The decimation of the empty circles leads to the 1-d chain depicted immediately below the sawtooth array. The dispersion relation in this case is given by, 
\begin{equation} 
(E-\epsilon+2 t)~\xi(E,k) = 0
\label{saw-3}
\end{equation}
 with $ \xi(E,k) = (E-\epsilon)^5 - 4 t (E-\epsilon)^4 - t^2 (E-\epsilon)^3 + 12 t^3 (E-\epsilon)^2 - 2 t^4 ( E-\epsilon) - 4 t^5 - 2 t^2 [(E-\epsilon)^3 - t (E-\epsilon)^2 - 3 t^2 (E-\epsilon) + 2 t^3] \cos~ka $. The non-dispersive, FB is seen to appear at $E=\epsilon-2t$, and is displayed in Fig.~\ref{fig-ek}(f), along with the dispersive bands. This FB energy is later extracted through a simple geometric construction on a finite size SGF, and the self similarity of the lattice is exploited to formulate the prescription of extracting a whole hierarchy of them as the SGF tends to its thermodynamic limit.

We have verified the band structures presented here, as obtained by the decimation scheme, with those obtained from a direct diagonalization of the Hamiltonian in each of the three cases discussed above. The kernels of the Hamiltonian are given in the Appendix, in Eq.~\eqref{ham-1(a)}, Eq.~\eqref{ham-1(c)}, Eq.~\eqref{ham-1(e)} respectively. The results match exactly.

\begin{figure}[ht]
\centering
(a)\includegraphics[width=.85\columnwidth]{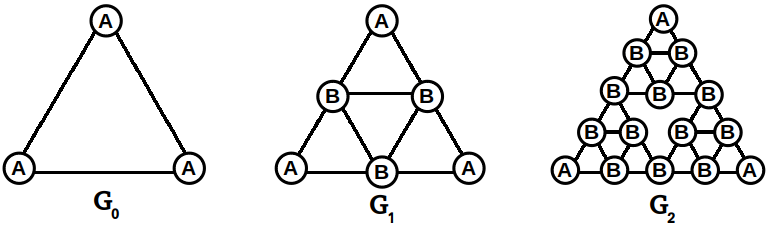}
(b)\includegraphics[width=.85\columnwidth]{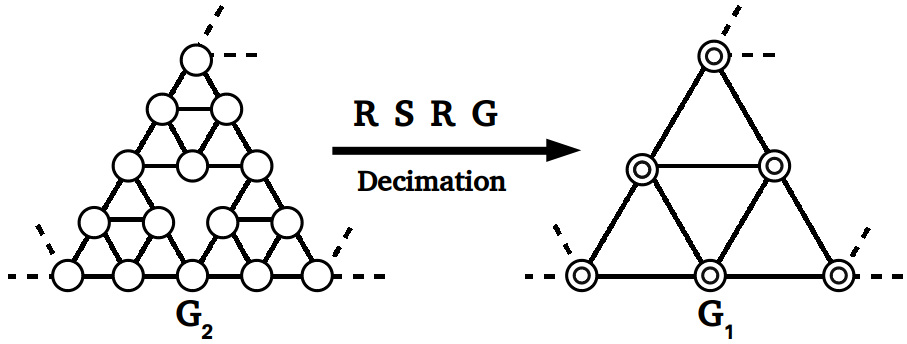}

\caption{(Color online) (a) The first three generations $G_0, G_1, G_2$ of a Sierpinski gasket fractal (SGF)  network. For a finite sized gasket the edge sites with a coordination number two are designated as $A$ and the bulk sites, having four nearest neighbors, are named $B$. (b) A second generation ($G_2$) cluster in an infinite SGF transforming into a first generation ($G_1$) cluster via an RSRG decimation scheme. The vertices surviving the decimation are marked by concentric  circles.}  
\label{sgf}
\end{figure}

\section{The Sierpinski gasket: engineering compact localization and the flat bands}
To understand and appreciate the method of constructing the CLS on an SGF substrate, and to work out the dispersion relations for a sawtooth array network consisting of finite size SGF motifs, we revisit the recursive method of generating an SGF geometry. The growth of the first ($G_1$) and the second ($G_2$) generation SGF, from the `seed' ($G_0$) are shown in Fig.~\ref{sgf}(a). A finite sized gasket has two kinds of sites, distinguished by the number of nearest neighbors they have. These are the `edge' sites, marked $A$, having a coordination number two,  and the `bulk' sites marked $B$, with a coordination number four, in Fig.~\ref{sgf}(a). Fig.~\ref{sgf}(b) shows a decimation process implemented on an infinite SGF, where all sites are of type $B$ (no edge sites are visible in the thermodynamic limit). In our analysis, throughout the present paper, we will take all sites to have a uniform on-site potential, viz, $\epsilon_A=\epsilon_B=\epsilon$ and a uniform nearest neighbor hopping integral $t$.

\begin{figure}[ht]
\centering
\includegraphics[width=.85\columnwidth]{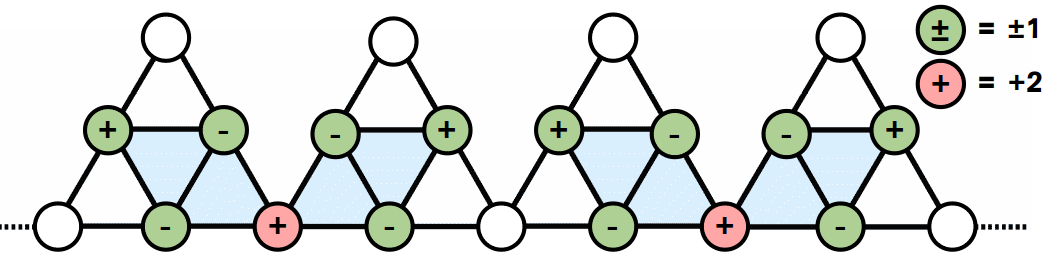}

\caption{(Color online) Distribution of  amplitudes on a first generation SGF array corresponding to the flat band energy $ E = -2 $ (non degenerate). The on-site potential for all the sites is set to zero, and the hopping integral is chosen as $ t = 1 $ along every arm of the gasket.}  
\label{gen1cls}
\end{figure}

\begin{figure}[ht]
\centering
(a)\includegraphics[width=.85\columnwidth]{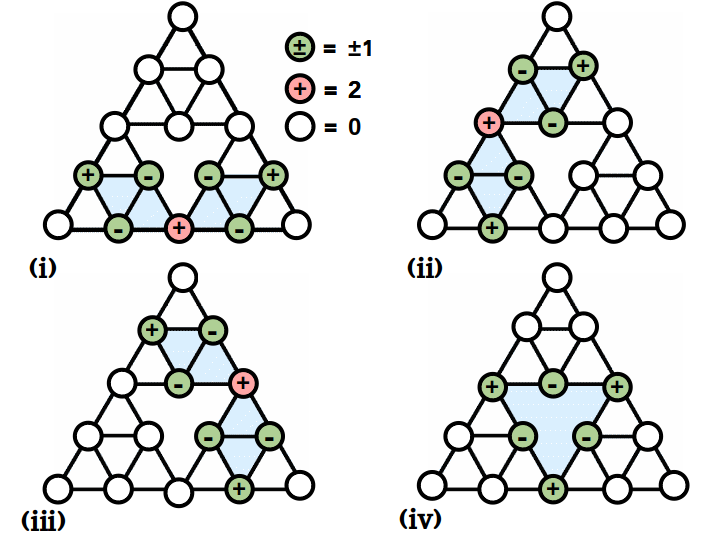}
(b)\includegraphics[width=.85\columnwidth]{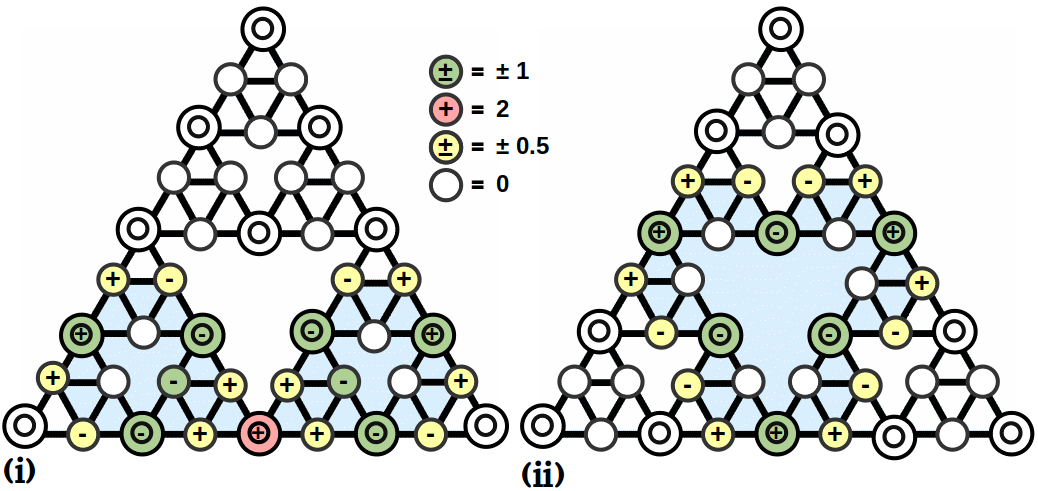}
\caption{(Color online) (a) Distribution of compact localized  amplitudes on a second generation ($G_2$) SGF network. The energy is $ E = -2 $ (four fold degenerate). The same energy and the same configuration of the amplitudes belong to a flat, non-dispersive band when we make a periodic array of $G_2$ units. Four different amplitude-distributions are shown in $(i)$, $(ii)$, $(iii)$, $(iv)$. (b) Distribution of  the CLS amplitudes on the third generation ($G_3$) SGF. The compact localization is there corresponding to an energy eigenvalue $ E = 1 $. This also happens to be the flat band energy for an array of $G_3$ motifs. Only two different amplitude-distributions are shown in (i) and (ii). The on-site potential for all the sites is set to zero, and the uniform nearest neighbor hopping integral is chosen as $t = 1$. }  
\label{gen22-31cls}
\end{figure}


It is already discussed in the last section, through Eq.~\eqref{saw-3} that, a flat band exists in a $G_1$-SGF sawtooth array at an energy $E = \epsilon-2 t$. This is because, we have set $\epsilon_A=\epsilon_B=\epsilon$. A generalization of this expression will be written down immediately, when we deal with larger versions of the basic SGF motif. The distribution of the amplitudes of the wavefunction for this particular CLS on a $G_1$-SGF sawtooth array is shown in Fig.~\ref{gen1cls}.


\begin{figure}[ht]
\centering
(a)\includegraphics[width=.8\columnwidth]{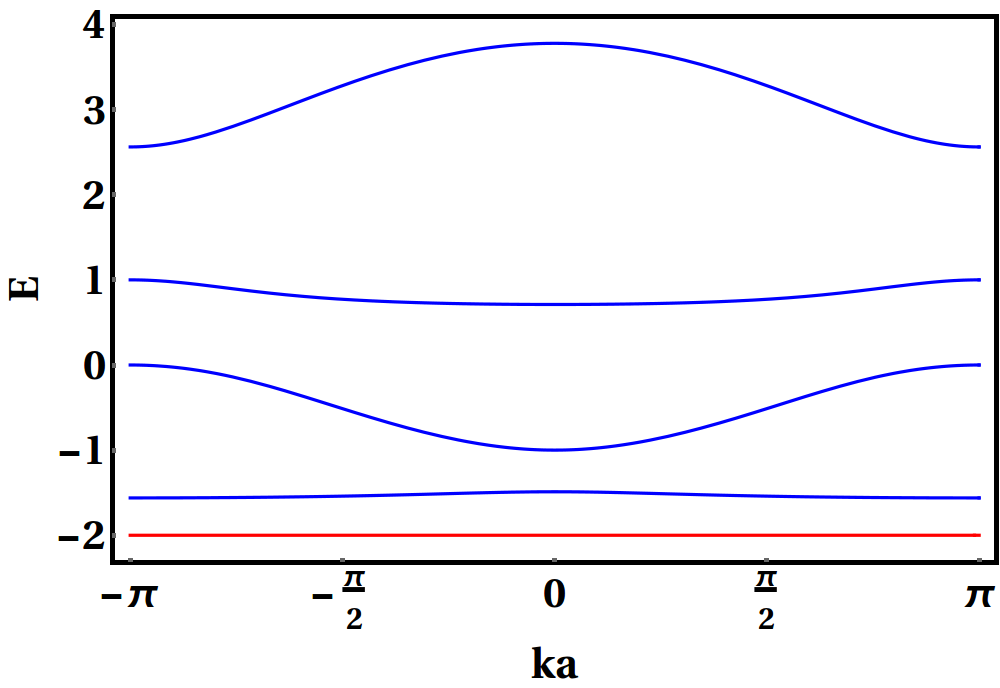}
(b)\includegraphics[width=.8\columnwidth]{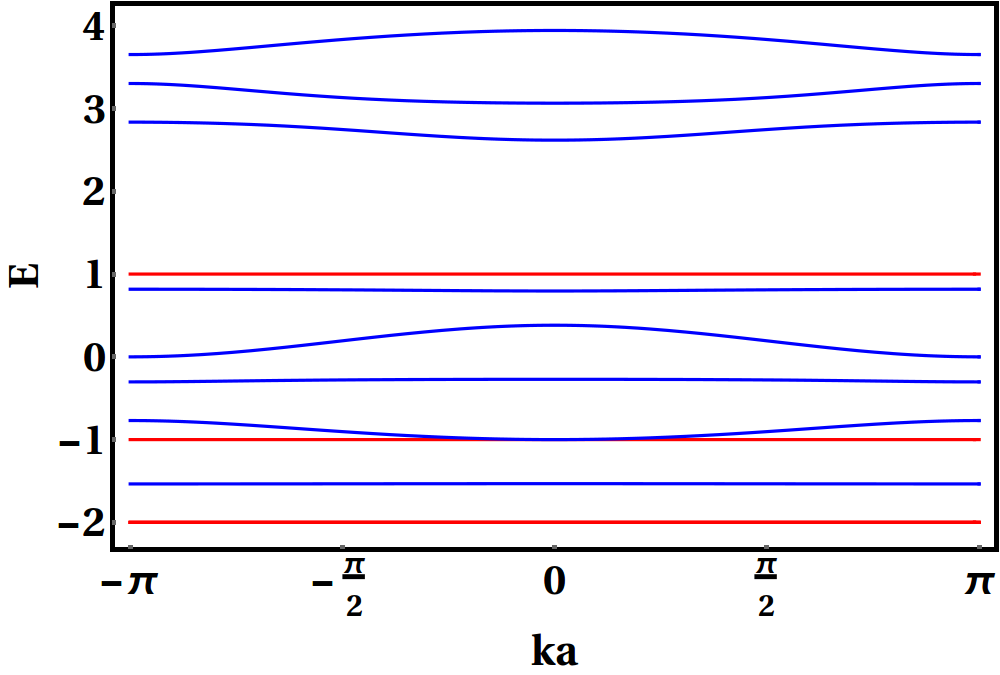}
(c)\includegraphics[width=.8\columnwidth]{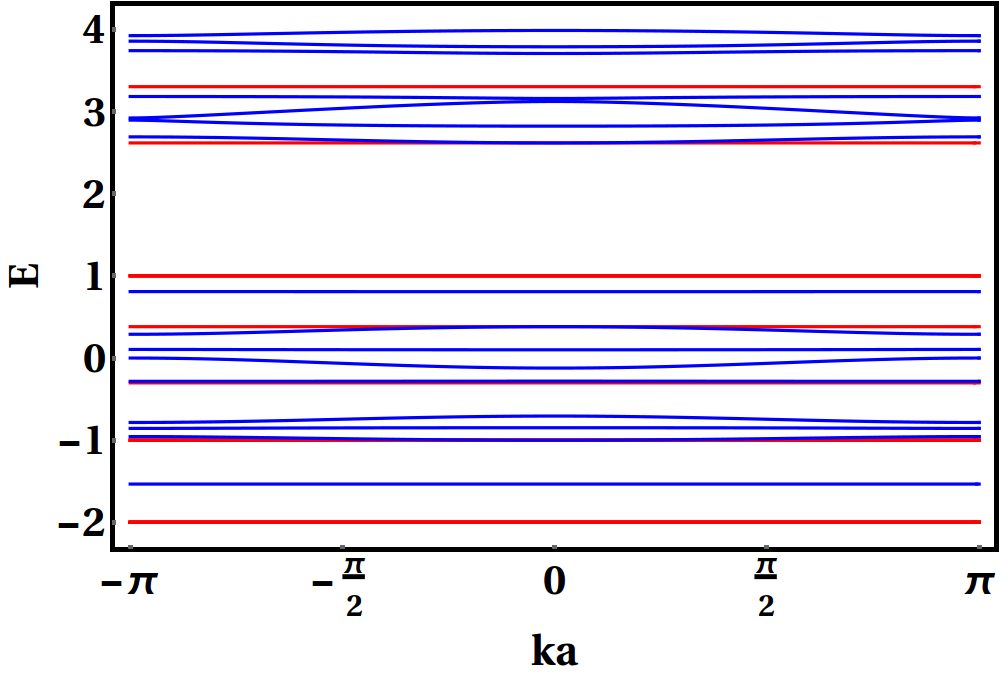}

\caption{(Color online) Energy-wave vector ($E$ - $ka$) dispersion relation of periodic arrays of (a) first, (b) second and (c) third generation Sierpinski gasket fractal (SGF) clusters as unit cells. We have chosen all onsite potentials $ \epsilon = 0 $ and hopping parameters $ t = 1$. Flat bands are drawn in red.}
\label{ek}
\end{figure}

It is to be noted that, this particular CLS can not be drawn on an isolated SGF at its $G_1$ stage, owing to the geometrical frustration built in the network. However, an array of $G_1$-SGF clusters circumvents this obstacle and compact `islands' with non-zero amplitudes of the wavefunction can be constructed, which rest on the backbone of the sawtooth array. Such islands are separated from their neighboring islands by sites where the amplitude of the wavefunction vanishes, an essential signature of the CLS.

The decimation scheme immediately opens up the  road to excavate the  CLS energies for isolated gaskets, or equivalently, the FB energy eigenvalues for arrays with unit cells comprising {\it any arbitrarily large} generation of an SGF. It should be appreciated that, when a sawtooth-like array of finite size gaskets are formed, the on-site potential $\epsilon_A$ of the `top' vertex of the triangle  renormalizes differently compared to the bulk sites having a on-site potential $\epsilon_B$ as one recursively uses the RSRG decimation scheme. We can make use of the RSRG recursion relations, that are already obtained earlier in the literature~\cite{arunava}, viz, 
\begin{eqnarray}
\epsilon_{A,n} & = & \epsilon_{A,n-1} + \frac{2 t_{n-1}^2 (E-\epsilon_{B,n-1})}{\Delta_{n-1}} \nonumber \\
\epsilon_{B,n} & = & \epsilon_{B,n-1} + \frac{4 t_{n-1}^2 (E-\epsilon_{B,n-1})}{\Delta_{n-1}} \nonumber \\
t_n & = & \frac{t_{n-1}^2 (E-\epsilon_{B,n-1}+2 t_{n-1})}{\Delta_{n-1}}
\label{rgeqn}
\end{eqnarray}
where, $\Delta_{n-1} = (E-\epsilon_{B,n-1})(E-\epsilon_{B,n-1}-t_{n-1})-2t_{n-1}^2$, to accomplish the task of extracting the CLS (or equivalently, the FB) energy eigenvalues. Here, $n$ denotes the number of times the RSRG decimation is implemented. Naturally, $n=0$ implies the basic, un-renormalized gasket. The scheme is outlined below.

Following the same line of analysis, as discussed with reference to Fig.~\ref{fig-ek} (e), an SGF array comprising any large $G_{n+1}$ motif as unit cell is reduced to an effective array of $G_1$ cells by renormalizing the SGF cluster $n$ times. As a result, the vertex (top site) now has an effective on-site potential $\epsilon_{A,n}$ and the atoms at the base and the bulk have on-site potentials $\epsilon_{B,n}$. We now follow the same algebra as we did before to work out the dispersion relation for the case shown in Fig.~\ref{fig-ek} (e). The dispersion relation for the linear chain of $G_{n+1}$ SGF clusters is now given by the equation,
\begin{equation}
(E-\epsilon_{B,n}+2 t_{n}) ~\chi_n (E,k) = 0
\label{disphigher}
\end{equation}
where, 
\begin{widetext}
\begin{equation}
\chi_n(E,k) = (E-\epsilon_{B,n}) (E-\epsilon_{B,n}-3 t_n) - 
\frac{4 t_n^4 (E-\epsilon_{B,n}+2 t_n)}{\Lambda_n} \cos^2(ka/2) -
2 t_n^2 \cos~ka
\label{dispfactor}
\end{equation}
\end{widetext}
Here, $\Lambda_n = (E-\epsilon_{A,n}) \Delta_n -2 t_n^2 (E-\epsilon_{B,n})$. Clearly, the CLS on an isolated gasket, as well as the FB energy eigenvalues are extracted from the zero's of the polynomial equation $E - \epsilon_{B,n} +2 t_n = 0$, and the dispersive bands for the `array' are obtained out of the equation $\chi_n (E,k)=0$.

The recursion relations in Eq.~\ref{rgeqn} has the unique property of recurring roots, in the sense that the roots of the equation $E-\epsilon_{B,n}+2t_n=0$ are again found among the roots of the equation $E-\epsilon_{B,n+1}+2t_{n+1}=0$ ~\cite{arunava}. Writing down explicitly, using Eq.~\eqref{rgeqn}, we find that the equation $E-\epsilon_{B,n+1}+2 t_{n+1}=0$ simplifies to,
\begin{equation}
(E-\epsilon_{B,n}+2 t_n) \left [(E-\epsilon_{B,n} )
(E-\epsilon_{B,n}-3 t_n) + 2t_n^2\right ] = 0
\label{nesting}
\end{equation}
The re-emergence of the roots at the $n$-th level in the $n+1$-th level is clear. Of course, there will be other roots as well, distinct from the previous ones.

In Fig.~\ref{gen22-31cls} we show how the amplitudes of the CLS wavefunctions are distributed among the sites of a finite size $G_2$ (in (a)) and $G_3$ (in (b)) SGF lattice. It should be quite clear now that, the same distributions will remain intact if we arrange such finite clusters periodically to design a decorated sawtooth chain of SGF's. Fig.~\ref{gen22-31cls}(a) shows three equivalent amplitude distribution of the four-fold degenerate CLS at $E=-2$ in $(i)$, $(ii)$ and $(iii)$. We have set $\epsilon_A = \epsilon_B=0$ and $t=1$.  The distributions are rotated with respect to each other by an angle of $2 \pi/3$. The fourth in the panel (a) shows another degenerate configuration where the amplitudes are pinned along the perimeter of the central blue shaded hexagon. This is kind of a {\it ring localized state}. As every `void' built in the SGF makes the neighboring sites sitting actually on an edge, the state depicted in Fig.~\ref{gen22-31cls}$a (iv)$ may easily be identified as an {\it edge state} in an SGF. In a sense, the three previous CLS distributions, occupying the vertices of the juxtaposed rhombii (shaded blue) also form a subset of the edge states, as the rhombii rest on any one of the sides (an edge again) of the full SGF cluster.
\begin{figure}[ht]
\centering
\includegraphics[width=.85\columnwidth]{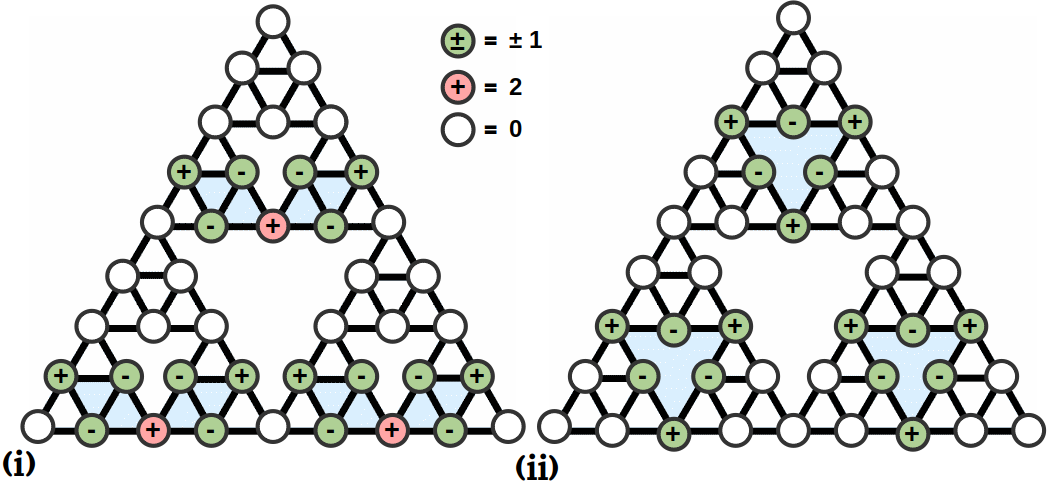}
\caption{(Color online) Spreading of two different variants of the CLS on the third SGF network corresponding to a flat band energy 
$E = -2$. The on-site potential for all the sites is set to zero, and the nearest neighbor hopping integral is chosen as $t = 1$. }  
\label{gen32cls}
\end{figure}
\begin{figure}[ht]
\centering
(a)\includegraphics[width=.4\columnwidth]{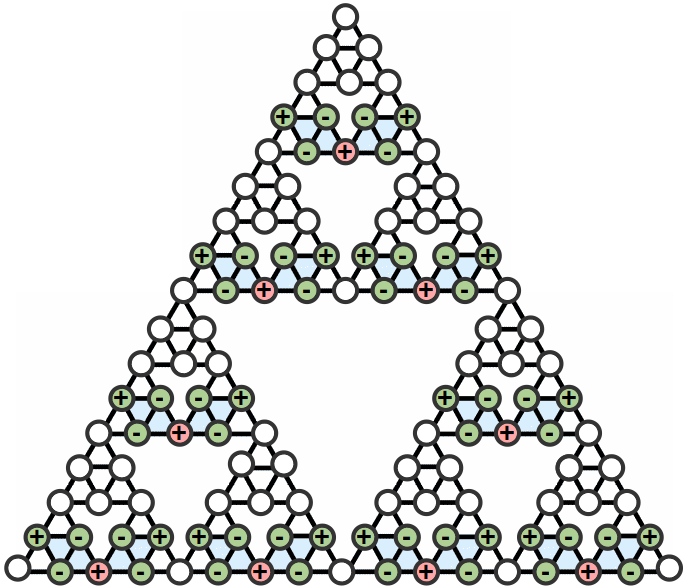}
(b)\includegraphics[width=.4\columnwidth]{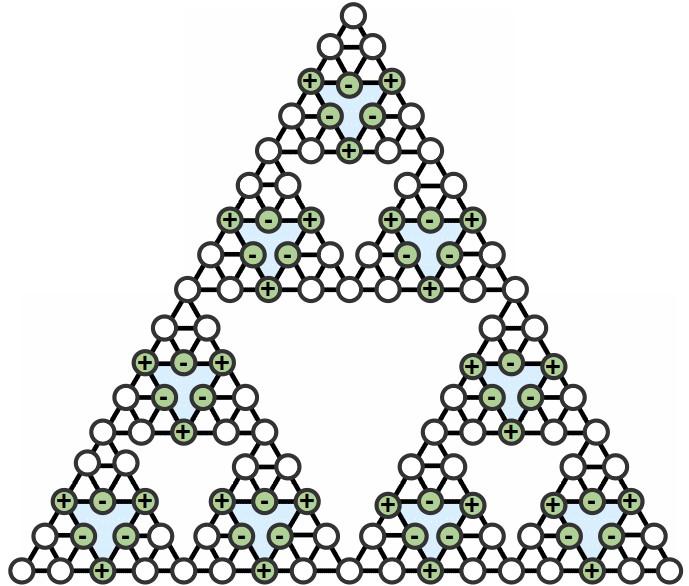}
(c)\includegraphics[width=.4\columnwidth]{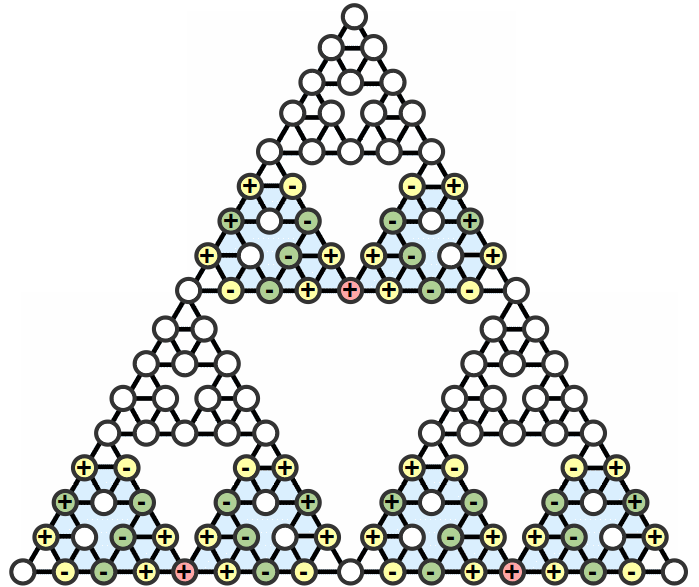}
(d)\includegraphics[width=.4\columnwidth]{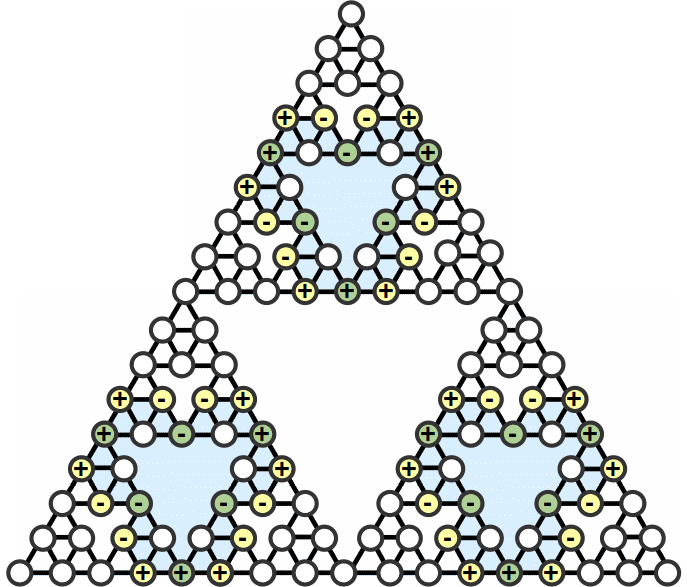}
\caption{(Color online) (a,b) Spreading of the CLS on a fourth generation SGF network corresponding to the flat band energy $ E = -2 $, and (c,d)  $ E = 1 $ respectively. The on-site potential for all the sites is taken to be zero, and the hopping integral is again chosen as $t = 1$. All colored sites indicate non zero amplitude, already discussed, and are expected to be {\it lit up}  in a photonic experiment. The color codes are the same as in Fig.~\ref{gen32cls} above.}
\label{gen4cls}
\end{figure}

To explore the edge state configurations at a different energy eigenvalue, we go over to a 3rd generation ($G_3$) SGF unit cell, and work out a CLS amplitude distribution for $E=1$, obtained as a solution of the equation $E-\epsilon_{B,1} + 2 t_1 = 0$. The distribution is shown in panels $(i)$ and $(ii)$ of Fig.~\ref{gen22-31cls}(b). It is to be observed that, even on this larger version of the gasket, the compactness of the states is preserved. The non-zero amplitudes now penetrate deeper into the bulk of the system, but the states retain their `edge' or ring localization character at this scale of length. 

To verify that the CLS states shown in Fig.~\ref{gen22-31cls} indeed belong to the FB's for an array of higher generation SGF motifs, in Fig.~\ref{ek} we plot the dispersion curves of linear arrays of finite size SGF with $G_1$, $G_2$ and $G_3$ clusters as unit cells. The red colored lines mark the flat bands, and the corresponding energy eigenvalues turn out to be localized on a compact, finite subcluster of the SGF. The distribution is preserved even if the finite size of the gasket is arbitrarily large. 

Interestingly, almost exact replicas of the amplitude patterns discussed above, have been experimentally observed recently in photonic lattices~\cite{xie,hanafi} where, a variant of the SGF structure reported here has been studied in two dimensions. In spite of the variation in construction, the ring-like CLS and the armchair patterns described here in Fig.~\ref{gen22-31cls}, are clearly seen. Even, our results collaborate an earlier theoretical work on flatbands in an SGF decorated 2D structure~\cite{kush} nicely. However, in none of the works referred to above, the FB eigenvalues and the CLS patterns are described for higher generation SGF's and we precisely stress upon this issue in this work here.

Before ending this section, it is important to pay attention to the following pertinent points.

$(i)$ The eigenvalues obtained from the equation $E-\epsilon_{B,n} + 2 t_n=0$ give compact {\it localized} states. Thus we have been able to work out a scheme through which at least an arbitrarily large subset of eigenvalues corresponding to the localized states in an SGF can be known {\it exactly}. The eigenvalues obtained at any $n$-th stage of the fractal are also the eigenvalues of the $n+1$-th, and naturally, of all subsequent stages. This RSRG method therefore provides a tool to locate an infinite number of energy eigenvalues as the fractal grows to an infinite size even. This is non-trivial, and is not attainable (at least, easily) by a direct diagonalization of arbitrarily large Hamiltonian matrices, as the fragmented Cantor-set like character of the eigenvalue spectrum can not ensure the exact {\it location} of a localized state on the energy scale.

$(ii)$ A look at the recursion relations in Eq.~\eqref{rgeqn} immediately reveals that setting $E=\epsilon_{B,n}-2 t_n$ makes the hopping at the next stage, viz, $t_{n+1}$ equal to {\it zero}. This is the signature of the corresponding eigenstates being localized in character. But, the point of interest is that, the eigenvalues extracted out of the equation $E-\epsilon_{B,n} + 2 t_n=0$ will keep the hopping integrals $t_1$, $t_2$, $t_3$ $.....$$t_n$ non-zero for all the previous RSRG decimation steps, from $1$ up to $n$. Thus localization will set in only at the $n+1$-th stage of  an SGF for the above energy eigenvalues, i.e. will be {\it delayed} in real space. The amplitudes of the wavefunction will have non-trivial distributions, penetrating deeper into the lattice, but eventually will get {\it frozen} over a cluster, forming the CLS. The size of the cluster will of course, depend on $n$, the stage of RSRG at which the CLS energies are extracted. In all the above examples we discussed a few priliminary cases.

$(iii)$ Apart from the CLS, the energy eigenvalues obtained out of the equation $E-\epsilon_{B,n}=0$ give rise to {\it extended}, non-Bloch type eigenstates. For an infinitely large SGF we encounter an infinite number of such isolated but densely packed extended states. The details have already been discussed in the literature~\cite{wang,arunava}.

\section{Conclusions}

We have tried to explain how the use of a real space renormalization group decimation technique can help extracting a countable infinity of compact localized states of a deterministic fractal network such as the Sierpinski gasket. The easiest construction is done exploiting the geometry of the lowest generation motif of the fractal, and the renormalization group recursion relations are then used to work out the subset of localized state-eigenvalues for bigger fractal substrates. These compact localized states are seen to show up at a scale of length that is determined by the stage of renormalization, $n$. Thus the scheme serves the dual purpose of extracting exact numerical values of the energy for which states are localized over finite areas on such a fractal, as well as in controlling its spatial distribution. The CLS found here are also the energies corresponding to the non-dispersive bands when one studies a periodic array of the fractal clusters.

\begin{figure}[ht]
\centering
(a)\includegraphics[width=.4\columnwidth]{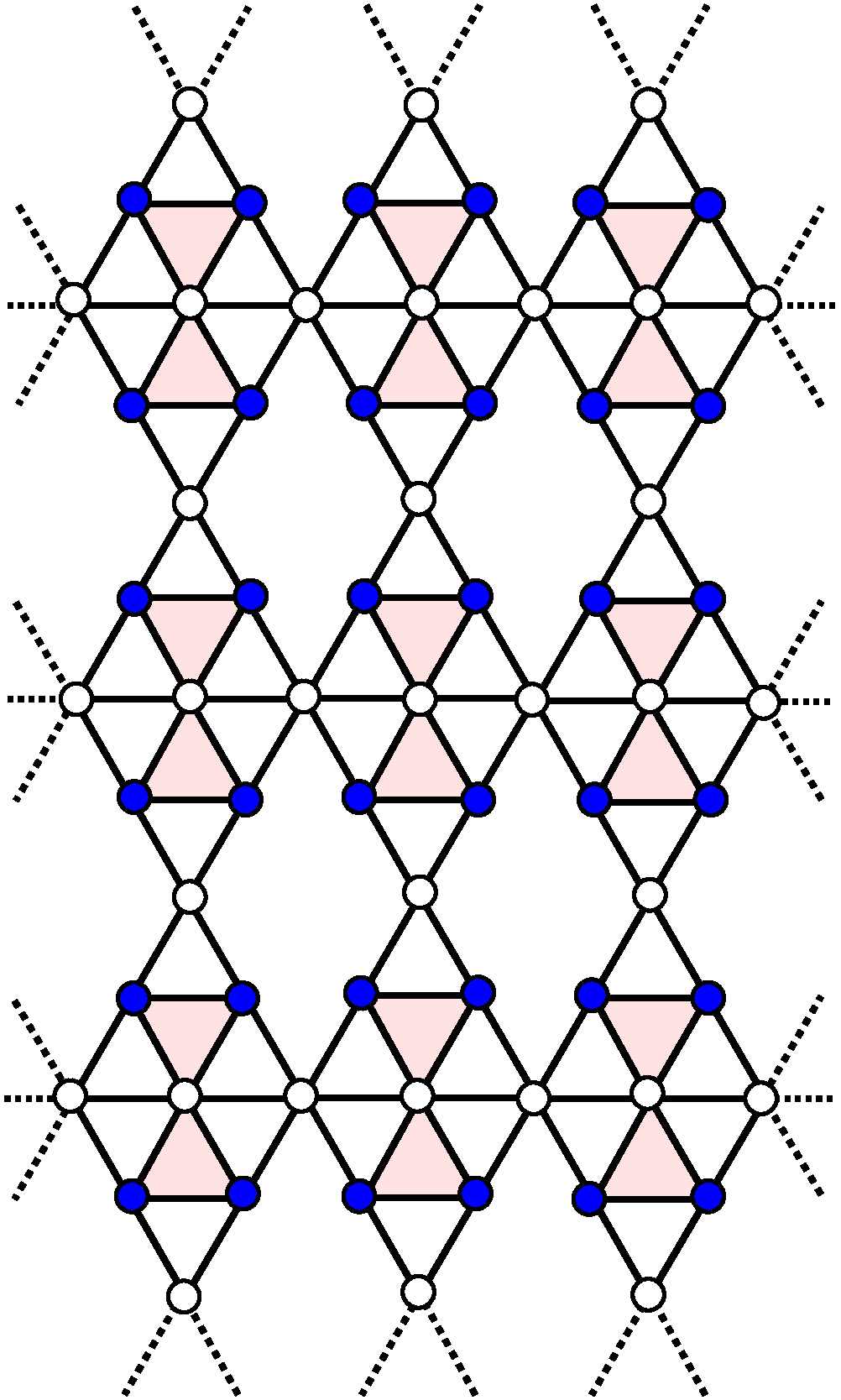}
(b)\includegraphics[width=.4\columnwidth]{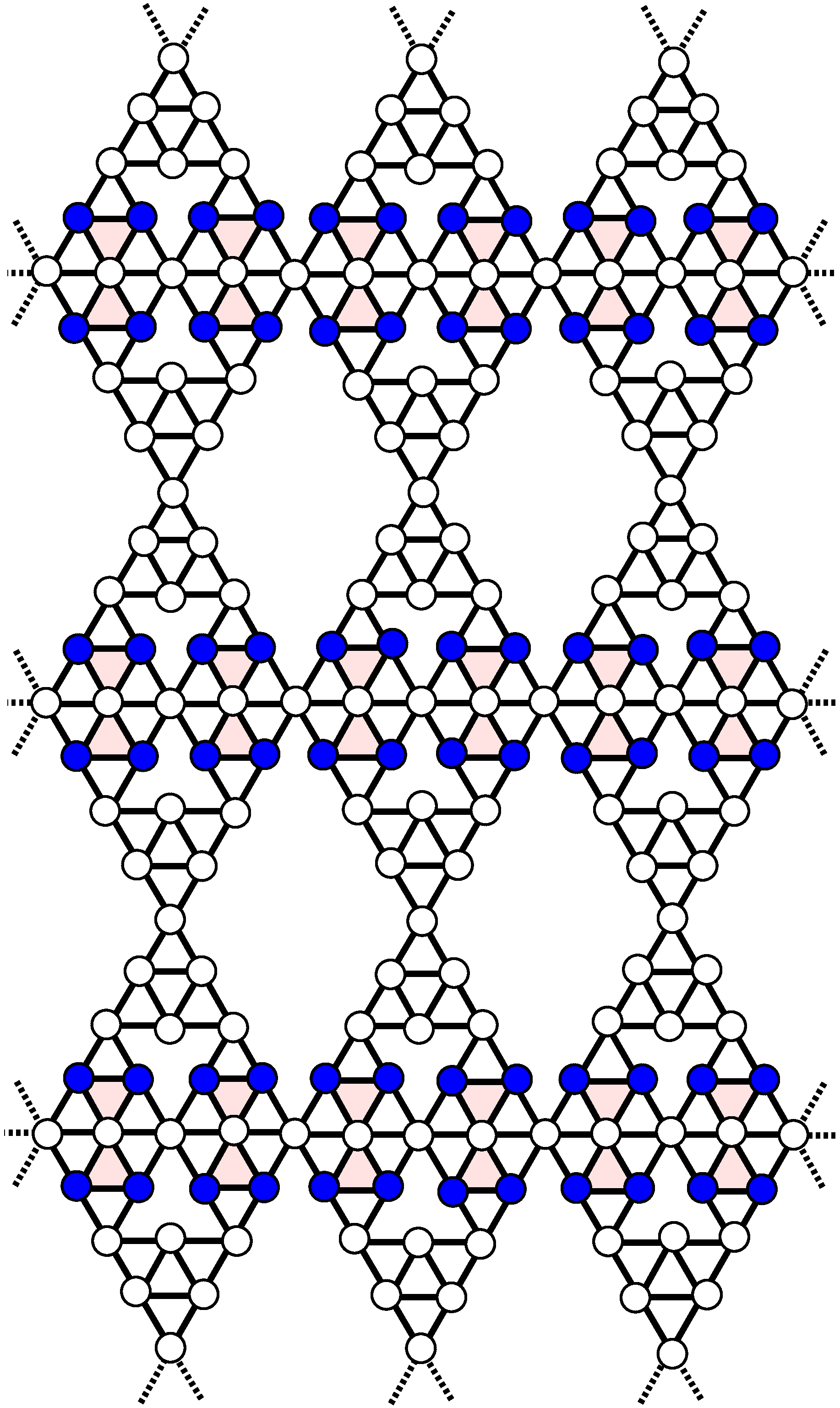}
\caption{(Color online) Distribution of  amplitudes on a two dimensional Sierpinski gasket (SPG) fractal network corresponding Flat Band energy $E = -1$. A first generation, and a second generation SGF motif serve as the unit cells in (a) and (b)  respectively. The on-site potential for all the sites is set to zero, and the hopping parameters are chosen as $ t = 1 $.  The values of the  amplitudes are $ \pm 1 $ ( blue color circle) and all others are zero.}  
\label{spg2dcls1}
\end{figure}

\begin{figure}[ht]
\centering
(a)\includegraphics[width=.4\columnwidth]{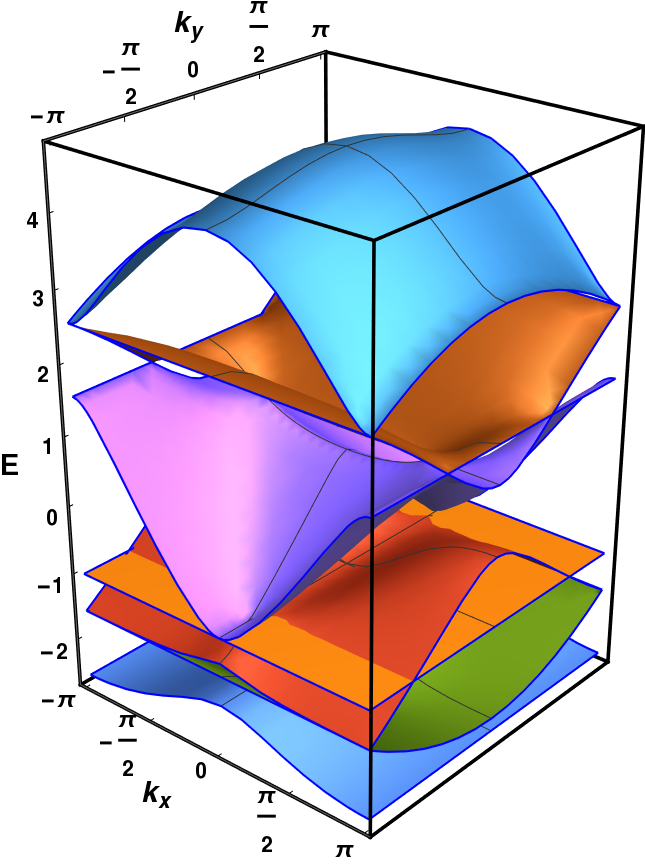}
(b)\includegraphics[width=.4\columnwidth]{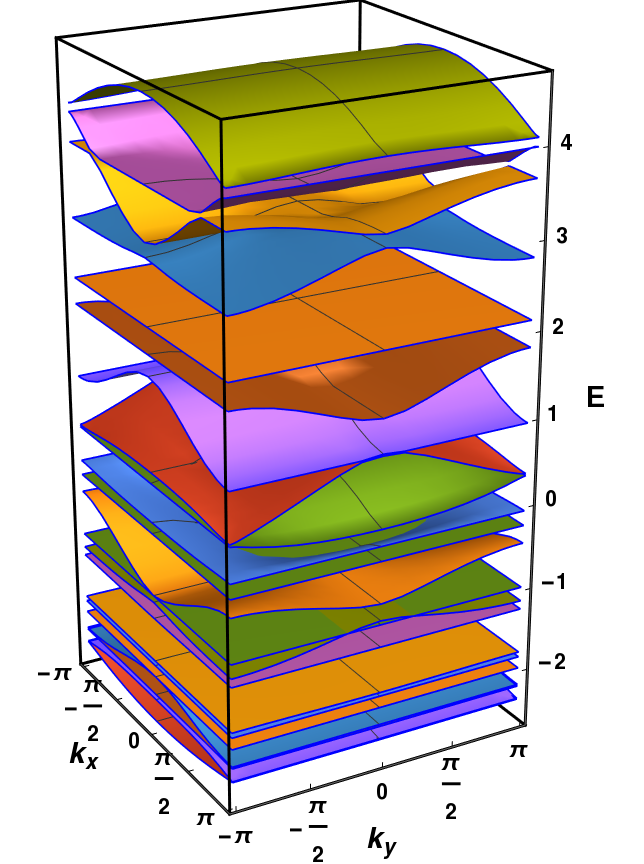}

\caption{(Color online) Energy-wave vector (E vs. $k_{x}$, $k_{y}$) dispersion relation of a two dimensional SGF network for (a) first, and (b) second generation respectively. We have chosen all onsite potentials $ \epsilon = 0 $ and hopping parameters $ t = 1 $.}  
\label{ek2d}
\end{figure}

The scheme outlined here can of course be extended to two dimensional versions of the lattice, such as the one shown in Fig.~\ref{spg2dcls1} (a) and (b). The dipersion relation, worked directly out of the Hamiltonian of the system is shown in Fig.~\ref{ek2d}. The dispersive, and the non-dispersive flat bands are clearly visible. Such distributions should be observable in experiments involving the evanescently coupled waveguides or ultracold atoms.

The renormalization decimation scheme can be implemented here to get a plethora of compact localized eigenstates and their distribution. However, the parameter space for two dimensional network is of a higher dimension than the two dimensional parameter space $(\epsilon, t)$ discussed in this work. Preliminary results have already been obtained in this direction and the rest of the results will be reported elsewhere.



\section{Acknowledgments}
 S. B. is thankful to Government of West Bengal for the SVMCM Scholarship. S. B. would like to  acknowledge Dr. Amrita Mukherjee  for  useful  discussions.

\newpage
\begin{widetext}
\section{APPENDIX}
The kernel of the Hamiltonian for the unit cell $\hat{\mathcal{H}}_{1(a)}(k) $ of a simple triangular array with two different types of hopping amplitudes $\Gamma$ and $t$ as shown in Fig.~\ref{fig-ek}(a) is given by,  

\begin{equation}
\hat{\mathcal{H}}_{1(a)}(k) = \left[ \begin{array}{cccccccccccccccc}
\epsilon + \Gamma e^{ika} +\Gamma e^{-ika} & t + t e^{-ika}\\
 t + t e^{ika} & \epsilon 
\end{array}
\right ] 
\label{ham-1(a)}
\end{equation}
The Hamiltonian corresponding to the unit cell of the `extended' sawtooth geometry, where there is an extra atomic site on each arm of the basic triangle (Fig.~\ref{fig-ek}(c) is:

\begin{equation}
\hat{\mathcal{H}}_{1(c)}(k) = \left[ \begin{array}{cccccccccccccccc}
\epsilon & t & 0 & t + t e^{-ika} & t e^{-ika} \\
t & \epsilon & t & 0 & 0 \\
0 & t & \epsilon & 0 & t \\
t + t e^{ika} & 0 & 0 & \epsilon & 0 \\
t e^{ika} & 0 & t & 0 & \epsilon
\end{array}
\right ] 
\label{ham-1(c)}
\end{equation}
and, the same for the triangular array with a 1st generation SGF motif as the unit cell (Fig.~\ref{fig-ek}(e)) is given by, 
 
\begin{equation}
\hat{\mathcal{H}}_{1(e)}(k) = \left[ \begin{array}{cccccccccccccccc}
\epsilon & t & 0 & t + t e^{-ika} & t e^{-ika} \\
t & \epsilon & t & t & t \\
0 & t & \epsilon & 0 & t \\
t + t e^{ika} & t & 0 & \epsilon & t \\
t e^{ika} & t & t & t & \epsilon
\end{array}
\right ] 
\label{ham-1(e)}
\end{equation}
For a linear array of unit cells comprising a second generation SGF clusters, the Hamiltonian is much more involved, and is given by,
 
\begin{equation}
\hat{\mathcal{H}}_{SPG(G2)}(k) = \left[ \begin{array}{cccccccccccccccc}
\epsilon & t & 0 & 0 & 0 & t & 0 & 0 & 0 & 0 & 0 & 0 & t e^{-ika} & t e^{-ika} \\
t & \epsilon & t & 0 & 0 & t & t & 0 & 0 & 0 & 0 & 0 & 0 & 0 \\
0 & t & \epsilon & t & 0 & 0 & t & t & 0 & 0 & 0 & 0 & 0 & 0 \\
0 & 0 & t & \epsilon  & t & 0 & 0 & t & t & 0 & 0 & 0 & 0 & 0 \\
0 & 0 & 0 & t & \epsilon & 0 & 0 & 0 & t & 0 & 0 & 0 & 0 & 0 \\
t & t & 0 & 0 & 0 & \epsilon & t & 0 & 0 & t & 0 & 0 & 0 & 0 \\
0 & t & t & 0 & 0 & t & \epsilon & 0 & 0 & t & 0 & 0 & 0 & 0 \\
0 & 0 & t & t & 0 & 0 & 0 & \epsilon & t & 0 & 0 & t & 0 & 0 \\
0 & 0 & 0 & t & t & 0 & 0 & t & \epsilon & 0 & 0 & t & 0 & 0 \\
0 & 0 & 0 & 0 & 0 & t & t & 0 & 0 & \epsilon & t & 0 & t & 0 \\
0 & 0 & 0 & 0 & 0 & 0 & 0 & 0 & 0 & t & \epsilon & t & t & t \\
0 & 0 & 0 & 0 & 0 & 0 & 0 & t & t & 0 & t & \epsilon & 0 & t \\
t e^{ika} & 0 & 0 & 0 & 0 & 0 & 0 & 0 & 0 & t & t & 0 & \epsilon & t \\
t e^{ika} & 0 & 0 & 0 & 0 & 0 & 0 & 0 & 0 & 0 & t & t & t &  \epsilon \\
\end{array}
\right ] 
\label{ham-spg2}
\end{equation}
while, for the two dimensional geometry depicted in Fig.~\ref{spg2dcls1}, the kernel of the Hamiltonian reads, 
\begin{equation}
\hat{\mathcal{H}}_{2d(8(a))}(k) = \left[ \begin{array}{cccccccccccccccc}
\epsilon & t & 0 & t e^{-ik_{y}a} & t e^{-ik_{y}a} & 0 & t \\
t & \epsilon & t & 0 & 0 & t & t \\
0 & t & \epsilon  & t & t e^{-ik_{x}a} & t+ te^{-ik_{x}a} & te^{-ik_{x}a}\\
te^{ik_{y}a} & 0 & t & \epsilon & t & t & 0 \\
te^{ik_{y}a} & 0 & te^{ik_{x}a} & t & \epsilon & t & 0\\
0 & t & t+te^{ik_{x}a} & t & t & \epsilon & t\\
t & t & te^{ik_{x}} & 0 & 0 & t & \epsilon\\
\end{array}
\right ] 
\label{ham-2d1}
\end{equation}
\end{widetext}


\begin{thebibliography}{99}
\bibitem{leykam} D. Leykam, A. Andreanov, and S. Flach, Adv. Phys. : X \textbf{3}, 1473052 (2018).
\bibitem{derzhko1} O. Derzhko, J. Richter and M. Maksymenko, Int. J. Mod. Phys. B \textbf{29}, 1530007 (2015).
\bibitem{bergholtz} E.J. Bergholtz and Z. Liu, Int. J. Mod. Phys. B \textbf{27}, 1330017 (2013). 
\bibitem{immanuel} I. Bloch, J. Dalibard, and J. Nsimbne, Quantum simulations with ultracold quantum gases, Nat. Phys. \textbf{8}, 267 (2012).
\bibitem{blatt} R. Blatt and C. F. Roos, Quantum simulation with trapped ions, Nat. Phys. \textbf{8}, 277 (2012).
\bibitem{rodrigo1} R. A. Vicencio and C. Mejía-Cortés, J. Opt. 16, 015706
(2014).
\bibitem{rojas} S. Rojas-Rojas, L. Morales-Inostroza, R. A. Vicencio, and A.
Delgado, Phys. Rev. A 96, 043803 (2017).
\bibitem{xia} S. Xia, Y. Hu, D. Song, Y. Zong, L. Tang, and Z. Chen, “Demonstration of flat-
band image transmission in optically induced Lieb photonic lattices,”Opt. Lett. \textbf{41}, 1435 (2016).
\bibitem{sutherland} B. Sutherland, Phys. Rev. B \textbf{34}, 5208 (1986).
\bibitem{gersen} H. Gersen, T. J. Karle, R. J. P. Engelen, W. Bogaerts, J. P. Korterik, N. F. van Hulst, T. F. Krauss, and L. Kuipers, “Real-space observation of ultraslow light in photonic crystal waveguides,” Phys. Rev. Lett. \textbf{94}, 073903 (2005).
\bibitem{seba} S. Mukherjee, A. Spracklen, D. Choudhury, N. Goldman, P. Öhberg, E. Andersson, and R. R. Thomson, “Observation
of a localized flat-band state in a photonic Lieb lattice,” Phys.
Rev. Lett. \textbf{114}, 245504 (2015).
\bibitem{rodrigo2} R. A. Vicencio, C. Cantillano, L. Morales-Inostroza, B. Real, C. Mejı́a-Cortés, S. Weimann, A. Szameit, and M. I. Molina, “Observation of localized states in Lieb photonic lattices,” Phys. Rev. Lett. \textbf{114}, 245503 (2015).
\bibitem{zong} Y. Zong, S. Xia, L. Tang, D. Song, Y. Hu, Y. Pei, J. Su, Y. Li, and Z. Chen,
“Observation of localized flat-band states in Kagome photonic lattices,” Opt. Express \textbf{24}, 8877–8885 (2016).
\bibitem{travkin} E. Travkin, F. Diebel, and C. Denz, “Compact flat band states in optically induced flatland photonic lattices,” Appl. Phys. Lett. 111, 011104 (2017).
\bibitem{drost} R. Drost, T. Ojanen, A. Harju, and P. Liljeroth, “Topological states in engineered atomic lattices,” Nat. Phys. 13, 668–671 (2017).
\bibitem{slot} M. R. Slot, T. S. Gardenier, P. H. Jacobse, G. C. P. van Miert, S. N. Kempkes, S. J. M. Zevenhuizen, C. Morais Smith, D. Van-maekelbergh, and I. Swart, “Experimental realization and characterization of an electronic Lieb lattice,” Nat. Phys. 13, 672–676 (2017).
\bibitem{gardenier} T. S. Gardenier, J. J. van den Broeke, J. R. Moes, I. Swart,C. Delerue, M. R. Slot, C. Morais Smith, and D. Vanmaekelbergh, “p orbital flat band and Dirac cone in the electronic honeycomb lattice,” ACS Nano 14, 13638–13644 (2020).
\bibitem{alexander} A. A. Khajetoorians, D. Wegner, A. F. Otte, and  I. Swart, Creating designer quantum states of matter atom-by-atom, Nat. Rev. Phys. \textbf{1}, 703 (2019).
\bibitem{miyahara} S. Miyahara, K. Kubo, H. Ono, Y. Shimomura, and N. Furukawa, “Flat-bands on partial line graphs–systematic method
for generating flat-band lattice structures,” J. Phys. Soc. Japan
74, 1918–1921 (2005).
\bibitem{mati} M. Hyrkäs, V. Apaja, and M. Manninen, “Many-particle dynamics of bosons and fermions in quasi-one-dimensional flat-band lattices,” Phys. Rev. A 87, 023614 (2013).
\bibitem{morales} L. Morales-Inostroza and R. A. Vicencio, “Simple method to construct flat-band lattices,” Phys. Rev. A 94, 043831 (2016).
\bibitem{rontgen} M. Röntgen, 1 C. V. Morfonios, 1 and P. Schmelcher, Compact localized states and flat bands from local symmetry partitioning, Phys. Rev. B \textbf{97}, 035161 (2018).
\bibitem{shang} J. Shang, Y. Wang, M. Chen, J. Dai, X. Zhou, J. Kuttner, G. Hilt, X. Shao, J. M. Gottfried, and K. Wu, Assembling molecular Sierpiński triangle fractals, Nat. Chem. \textbf{7}, 389 (2015).
\bibitem{tait} S. L. Tait, Surface chemistry: Self-assembling Sierpiński triangles, Nat. Chem. \textbf{7}, 370 (2015).
\bibitem{sun} Q. Sun, L. Cai, H. Ma, C. Yuan, and W. Xu, On-surface
construction of a metal–organic Sierpiński triangle, Chem. Commun. \textbf{51}, 14164 (2015).
\bibitem{xie} Yuqing Xie, 1 Limin Song, Wenchao Yan, Shiqi Xia, Liqin Tang,  Jun-Won Rhim, and Zhigang Chen, "Fractal-like photonic lattices and localized states arising from singular and nonsingular flatbands", APL Photon. \textbf{6}, 116104 (2021); doi: 10.1063/5.0068032
\bibitem{hanafi} H. Hanafi, P. Menz, and C. Denz, "Localized states emerging from singular and non-singular flat bands in a frustrated fractal-like photonic lattice", Adv. Opt. Mat. \textbf{10}, 2102523 (2022). DOI: 10.1002/adom.202102523
 \bibitem{liu} C. Liu et al. Sierpiński structure and electronic topology in bi thin films on insb(111)b surfaces. Phys. Rev. Lett. \textbf{126}, 176102 (2021).
\bibitem{kempkes} S. N. Kempkes, M. R. Slot, S. E. Freeney, S. J. M. Zevenhuizen, D. Vanmaekelbergh, I. Swart, and C. M. Smith, Design and
characterization of electrons in a fractal geometry, Nat. Phys.
15, 127 (2019).
\bibitem{xu} X.-Y. Xu, X.-W. Wang, D.-Y. Chen, C. M. Smith, and X.-M. Jin, Shining light on quantum transport in fractal networks,
arXiv:2005.13385
\bibitem{yang} Z. Yang, E. Lustig, Y. Lumer, and M. Segev, Photonic Floquet topological insulators in a fractal lattice, Yang et al. Light: Science and Applications, \textbf{9}, 128 (2020). https://doi.org/10.1038/s41377-020-00354-z
\bibitem{shriya} Shriya Pai and Abhinav Prem, "Topological states on fractal lattices", 
Phys. Rev. B \textbf{100}, 155135 (2019).
\bibitem{fischer} S. Fischer, M. vanHooft, T. vander Meijden, C. MoraisSmith, L. Fritz, and M. Fremling,  Robustness of chiral edge modes in fractal-like lattices below two dimensions: a case study. Phys. Rev. Res. \textbf{3}, 043103 (2021).
\bibitem{fremling} M. Fremling, M. van Hooft, C. M. Smith, and L. Fritz, Existence of robust edge currents in Sierpiński fractals. Phys. Rev. Res. \textbf{2}, 013044 (2020).
\bibitem{manna} S. Manna, B. Pal, W. Wang, and A. E. B. Nielsen, Anyons and fractional quantum Hall effect in fractal dimensions. Phys. Rev. Res. \textbf{2}, 023401 (2020).
\bibitem{brezinska} M. Brzezińska, A. M. Cook, and T. Neupert, Topology in the Sierpiński-Hofstadter problem. Phys. Rev. B \textbf{98}, 205116 (2018).

\bibitem{domany} E. Domany, S. Alexander, D. Bensimon, and L. P. Kadanoff, Phys. Rev. B \textbf{28}, 3110 (1983).
\bibitem{banavar} J. R. Banavar, Leo Kadanoff, and A. M. M. Pruisken, Energy spectrum for a fractal lattice in a magnetic field, Phys. Rev. B \textbf{31}, 1388 (1985). 
\bibitem{wang} X. R. Wang, Phys. Rev. B 51, 9310 (1995).
\bibitem{arunava} A. Chakrabarti, Exact results for infinite and finite Sierpinski gasket fractals: extended electron states and transmission properties, J. Phys.: Condens. Matt. \textbf{8}, 10951 (1996). DOI 10.1088/0953-8984/8/50/021
\bibitem{anderson} P. W. Anderson, Absence of Diffusion in Certain Random Lattices, Phys. Rev. \textbf{109}, 1492 (1958).

\bibitem{derzhko2} O. Derzhko, J. Richter, A. Honecker, M. Maksymenko, and R. Moessner, Low-temperature properties of the Hubbard model on highly frustrated one-dimensional lattices, Phys. Rev. B i\textbf{81}, 014421 (2010).
\bibitem{schulze} M. Schulze, D. Bercioux, and D. F. Urban, Adiabatic pumping in the quasi-one-dimensional triangle lattice, Phys. Rev. B \textbf{87}, 024301 (2013).
\bibitem{cai} X. Cai, S. Chen, and Y. Wang, Quantum dynamics in driven sawtooth lattice under uniform magnetic field, Phys. Rev. A \textbf{87}, 013607 (2013).
\bibitem{magnus} M. Johansson, U. Naether, R. A. Vicencio, Compactification tuning for nonlinear localized modes in sawtooth lattices, Phys. Rev. E \textbf{92}, 032912 (2015).
\bibitem{boong} T. Zhang and Gyu-Boong Jo, One-dimensional sawtooth and zigzag lattices for ultracold atoms, Scientific Reports | 5:16044 | DOI: 10.1038/srep16044 (2015).
\bibitem{maimaiti} W. Maimaiti, A. Andreanov, H. C. Park,  O.  Gendelman, and S. Flach, Compact localized states and flat-band generators in one dimension, Phys. Rev. B \textbf{95}, 115135 (2017).
\bibitem{peotta} V. A. J. Pyykkonen, S. Peotta, P. Fabritius, 
J. Mohan, T. Esslinger, and P. Torma, Phys. Rev. B \textbf{103}, 144519 (2021).
\bibitem{kush} B. Pal and K. Saha, "Flat bands in fractal-like geometry", Phys. Rev. B \textbf{97}, 195101 (2018).
\end{thebibliography}
\end{document}